\documentclass[aps,prl,twocolumn,superscriptaddress]{revtex4}

\usepackage{graphicx}
\usepackage{bm}
\usepackage{amsmath}
\usepackage{amssymb}
\usepackage{hyperref}

\begin{document}
\title{Chirality-Selective Transport of Benzene Molecules on Carbon Nanotubes}

\author{Zhao Wang}
\email{zw@gxu.edu.cn}
\affiliation{Department of Physics, Guangxi University, Nanning 530004, China}

\begin{abstract}
Using molecular dynamics simulations, we predict an effect of chirality on the conduction of benzene molecules along the surface of carbon nanotubes (CNTs) subjected to a thermal gradient. The group drift velocity of the molecules is found to be maximal in the case of an armchair CNT, and to decrease with decreasing chiral angle. This chirality effect on thermodiffusion is induced by a variation in the optimized paths of molecules that change with different electronic overlap at the interface. The mechanism for the thermophoretic transport is identified to be coupled with a gradient of adsorbate-substrate interaction energy, which originates from the anharmonic nature of the van der Waals potential.
\end{abstract}

\maketitle

\section{Introduction}

Molecular transport via nanostructures holds great promise for a wide range of applications, including separation of chemicals, drug delivery, energy conversion and storage, and so forth.\cite{Sun2016,Li2015}. Among a large variety of nanomaterials, carbon nanotubes (CNTs) have drawn considerable attention for directional molecular conduction because of their chemical inertness and peculiar structure that are strongly coupled to their electronic properties. Water conduction through CNT channels was predicted by Hummer \textit{et al.} using molecular dynamics (MD) simulations,\cite{Hummer2001} and was afterward observed in experiments.\cite{Holt2006} Intensive efforts have subsequently been devoted to study the conduction of water and other liquids through CNTs.\cite{Goldsmith2009,Kalugin2008} Notably, water molecules confined inside CNTs exhibit directional motion under an applied temperature difference.\cite{Zambrano2009,Oyarzua2018} However, energy barriers exist for many molecules and ions to enter CNT channels due to the high aspect ratio of CNTs,\cite{Peter2005,Comer2015} whereas the outer surface of CNT could be more accessible.\cite{Russell2012}

For instance, Mao and Sinnott predicted a spiral diffusion path of nonspherical organic molecules in CNTs using MD and density functional theory (DFT) calculations.\cite{Mao2002} Barreiro \textit{et al.} observed directional motion of a gold nanoparticle on an outer wall of a multiwalled CNT driven by thermal gradients in scanning electron microscopy experiments after a theoretical prediction.\cite{Barreiro2008,Schoen2006} Recently, Becton and Wang utilized a thermal gradient on a graphene sheet to control the motion of a nanoflake physically adsorbed on it, and ascribed the origin of the driving force to the variation in the kinetic energy imparted to the nanoflake due to the interaction with the graphene sheet along the temperature gradient.\cite{Becton2014} However, Panizon \textit{et al.} reported a driving force independent of the local gradient magnitude on a gold cluster transmitted through scattering with flexural phonon waves on a graphene substrate.\cite{Panizon2017} Furthermore, Leng \textit{et al.} observed negative thermophoresis of a single-walled CNT nested inside of two separate outer CNTs held at different temperatures, revealing the mechanism of competition between the thermally induced edge force and the interlayer attraction force.\cite{Leng2016} The origin of the thermophoretic driving force for the molecular transport is under debate.

Toward a better understanding on the mechanism for the molecular transport on nanostructure surface, this paper studies the thermodiffusion of benzene molecules adsorbed on single-walled CNTs using MD. Benzene is chosen to be the adsorbate because of previously discovered features of $\pi$-$\pi$ stacking between $sp^{2}$ hybridized carbons.\cite{Bjork2010,Becton2014} The effect of the CNT structure on the molecular transport is investigated in different conditions with various molecular densities and temperatures.

\section{Method}

CNTs have the walls that are indistinguishable from graphene sheets rolled at specific and discrete chiral angles $\phi$. The structure of a CNT is determined by either the combination of $\phi$ and the radius $R$, or two integers $n$ and $m$ which define a circumferential vector in the hexagonal graphene lattice. We consider a periodic cell of an infinite free-standing single-walled CNT, its length is $p \ell_{0} \approx 20\;\mathrm{nm}$, where $p$ is an integer and $\ell_{0}$ is the chirality-dependent length of a CNT unit cell.\cite{CNTbook2001} 

In analogy to experiments, a heat flow is generated by defining a heat source and a heat sink at the both sides of the CNT at temperature $T_{1}$ and $T_{2}$, respectively.\cite{Barreiro2008,Zambrano2009,Oyarzua2018} The thickness of each thermal reservoir is about $0.6\;\mathrm{nm}$. The rate of the Nose-Hover thermostat is per $10\;\mathrm{fs}$. The boundaries along the tube axis are set periodic in order to ensure a steady flow of the molecules. A thin ($0.3\;\mathrm{nm}$) segment of atoms is set to be motionless near the periodic boundary, this rigid layer forces the heat to flow only in one direction. A test run was carried out on a (10,10) CNT to make sure that the trajectory of the molecules does not depend significantly on the size of this motionless parts, as shown in the Supporting Information (S1). 

\begin{figure}[htp]
\centerline{\includegraphics[width=9cm]{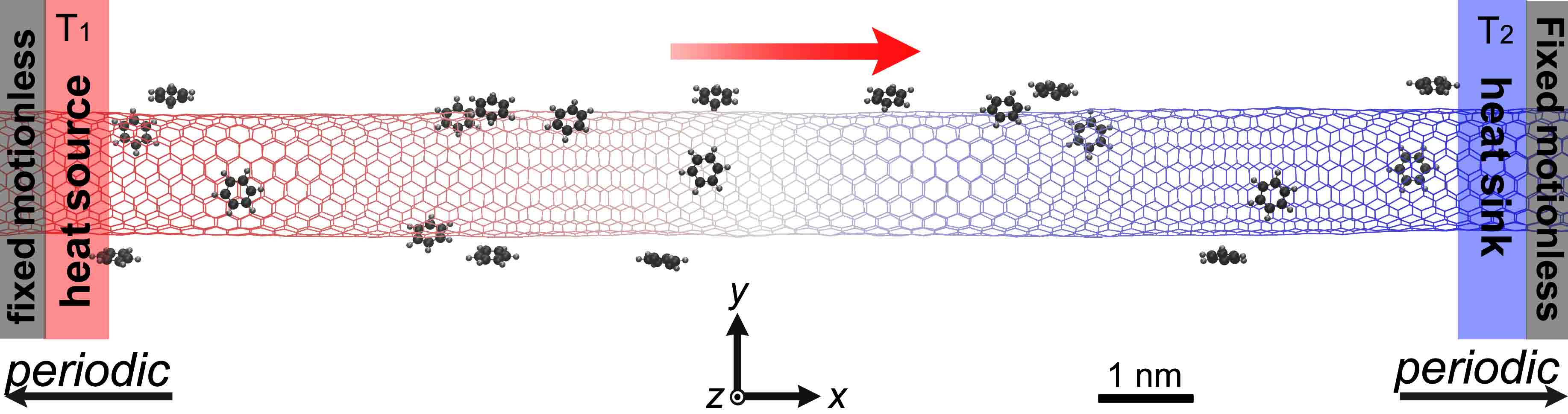}}
\caption{\label{F1}
Schematic of benzene molecules adsorbed on a segment of a (10,10) CNT, in which a temperature gradient is applied along the tube axis by defining two heat reservoirs at different temperatures at both sides.}
\end{figure}

The potential energy of the system is described by the adaptive interatomic reactive empirical bond order (AIREBO) potential, which is a collection of those of individual bonds mimicking many-body effects. It consists of both the covalent and the non-covalent interaction potentials. The covalent part is described by the reactive empirical bond order (REBO) potential,\cite{Brenner1990} while the non-covalent part is included by adding a long-range van der Waals (vdW) $6$-$12$ Lennard-Jones (LJ) force field. A cutoff radius of about $1.1\;\mathrm{nm}$ is used for the long-range interaction. The parameterization and benchmarks of this force field is provided elsewhere.\citep{Stuart2000} We note that the specific description to the effects of bond rotation and torsion in the AIREBO potential is important for simulating the adsorption process, owing to the deformation of the substrate caused by the adsorbates or thermal fluctuations.

Different sets of simulations are performed by varying CNT type, molecular density, temperature and thermal gradient using the simulation package Lammps.\cite{Plimpton95} A number of benzene molecules are physically adsorbed on the outer CNT surface as shown in Fig.\ref{F1}. The thermal reservoir regions are then controlled to progressively reach the set temperatures $T_{1}$ and $T_{2}$ by the canonical Nos\'{e}-Hoover thermostat at a time step of $1.0\;\mathrm{fs}$. A temperature gradient along the CNT is generated by the thermal resistance to the heat flow, and drives the adsorbates along the CNT surface.\cite{Yang2015,Lin2014} A box-indexing method is used to track the trajectories of the center of mass of the adsorbates in periodic condition as described in the Supporting Information (S2). Moreover, Lammps input scripts are provided in the Supporting Information (S5) for the ease of interested readers to reproduce the same calculations for checking if the results are correct.

\section{Results and Discussion}

\begin{figure}[htp]
\centerline{\includegraphics[width=9cm]{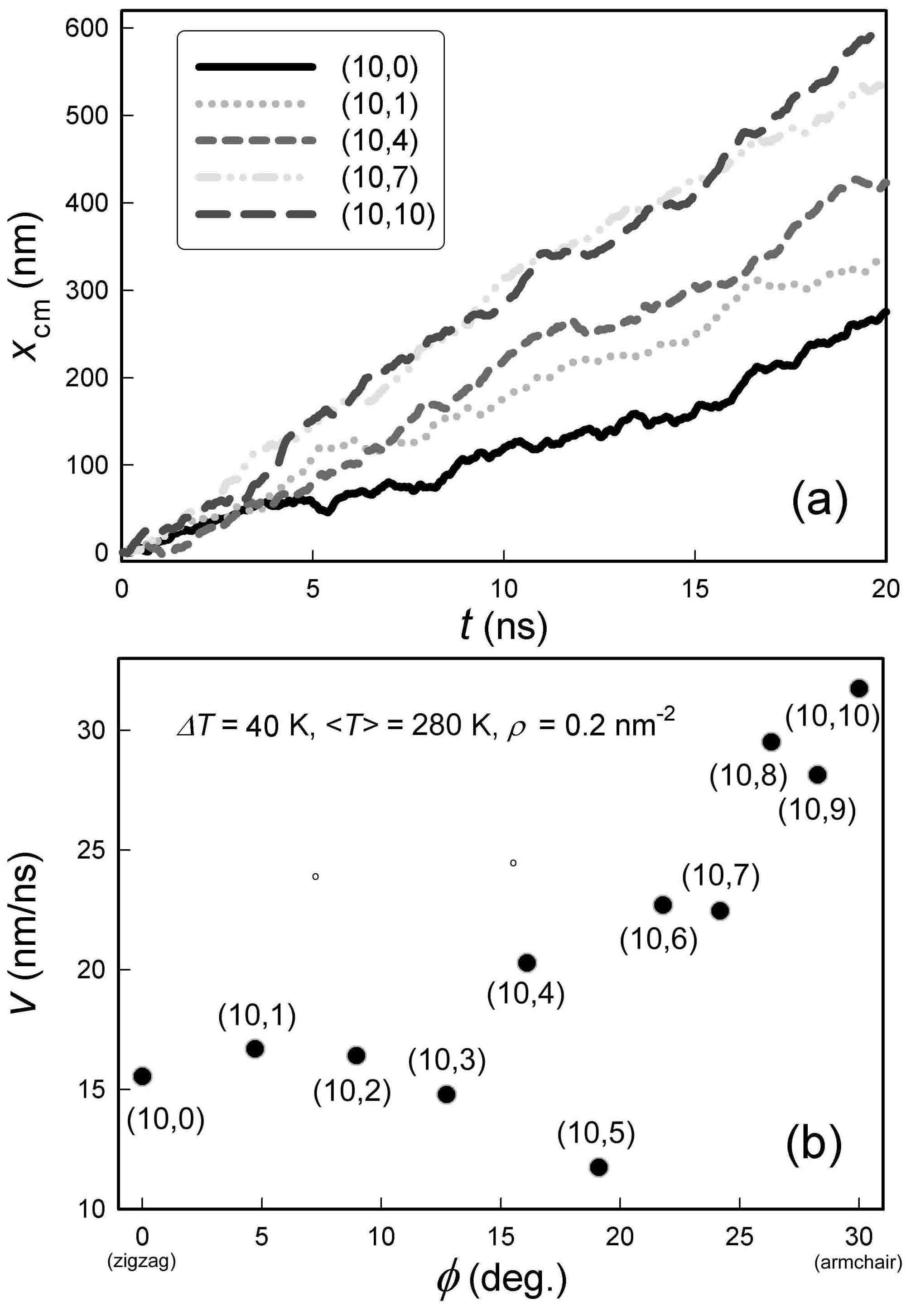}}
\caption{\label{F2}
(a) Position of the center of mass $x_{cm}$ of the benzene molecules along the tube axis vs the simulation time $t$, on CNTs of different chirality at a temperature difference between $T_{1}=260\;\mathrm{K}$ and $T_{2}=300\;\mathrm{K}$ with a constant number density of adsorbates per unit of CNT surface area $\rho=0.2\;\mathrm{nm^{-2}}$. (b) Drift velocity $v$ vs the chiral angle $\phi$ for eleven CNTs of $n=10$ and $m=0,1,2,...,10$.}
\end{figure}

Fig.\ref{F2} (a) shows the evolution of the $x$ position of the center of mass of the collection of the benzene molecules adsorbed on five CNTs of different chirality. High-speed molecular transport is observed: The benzene group can move dozens of nanometers per nanosecond. They travel the fastest on an armchair (10,10) CNT, and is slower on chiral and zigzag ones. The group drift velocity $v$, defined as the slope of the $x_{cm}-t$ curve, increases with increasing chiral index $m$ from the zigzag ($10$,$0$) to armchair ($10$,$10$) CNT. These curves are not straight lines because a major part of the motion of molecule is random in the femtosecond regime,\cite{Panizon2017} i.e., a longer data averaging time will make the curves straighter. Moreover, animations illustrating the molecular conduction are provided in the Supporting Information (S6-S9).

Fig.\ref{F2} (b) shows a comprehensive view of the drift velocity in relation to the chiral angle $\phi$ of eleven CNTs. It is seen that, with increasing $\phi$, $v$ remains roughly invariant until $\phi$ reaches $19.1^\circ$ for the ($10$,$5$) CNT. Then, $v$ rapidly increases and attains a maximum at $\phi=30^\circ$ for the armchair ($10$,$10$) CNT. According to the experimental observation of orientation-dependent interactions between CNTs and graphene,\cite{Chen2013} the dependence of the molecular transport of CNTs on their chirality should be strongly correlated with the registry configuration of $\pi$ orbitals at the interface as we are going to explain in details below.\cite{Kolmogorov2004,Zhao2015JPCL,Garnier2016JPCL} Note that $\phi = 19.1^\circ$ is a so-called ``magic'' chiral angle of CNT, at which a structural transition may occur at the CNT interface.\cite{Artyukhov2014}

\begin{figure}[htp]
\centerline{\includegraphics[width=9cm]{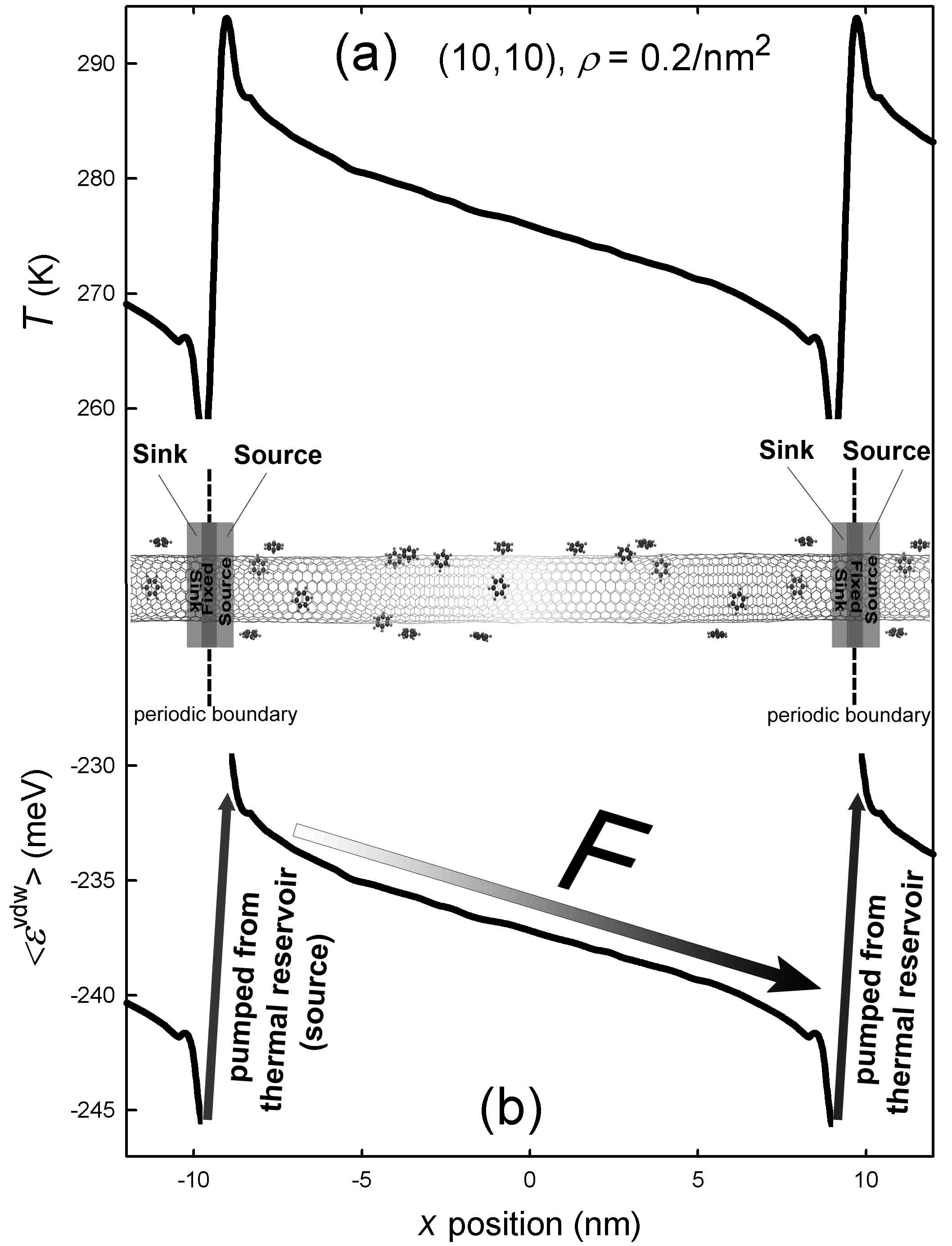}}
\caption{\label{F3}
Profiles of (a) temperature and (b) time-averaged vdW potential energy of the benzene-CNT interaction $\varepsilon^{vdW}$ along the CNT ($x$ axis). }
\end{figure}
 
Identification of the driving force is the first step to solve the puzzle of the chirality dependence of the molecular transport. As mentioned before, several possible origins of the driving force have been suggested in the literature for the molecular transport on CNTs or graphene subjected to a thermal gradient. These include differences in kinetic energy, flexural phonon waves, and a competition between thermally induced edge forces and interlayer attraction forces. Among these mechanisms, the edge force is not applicable to the periodic system simulated in this work.\cite{Leng2016}. Thermophoretic force due to difference in the kinetic energy is typical for thermodiffusion in gaseous or liquid phase,\cite{Becton2014} but it unlikely dominates the mass transport at gas-solid interfaces. The mechanism based on flexural phonon waves relies on the particular ZA phonons in graphene which have a very long mean-free-path.\cite{Panizon2017} This is based on a selection rule that stems from inversion symmetry for two-dimensional materials, and does not apply to one-dimensional nanostructures like CNTs.

So, where does the driving force come from? Here, the driving force is identified as stemming mainly from a gradient in the vdW potential. This gradient is induced by a discrepancy in the adsorbate-substrate distance due to the non-uniform temperature field, which arises from the anharmonicity of the interatomic potential. i.e. a molecule can be considered as an anharmonic oscillator anchored to the CNT surface, with an averable distance between benzene and graphene that increases with temperature. Thus, a larger distance at the hot side corresponds to an interaction potential higher than that at the cold side. The data supporting this mechanism are plotted in Fig.~\ref{F3}. This driving force is fundamental for molecular mass transport on nanostructures, and could be useful to explain results of previous experiments.\cite{Barreiro2008,Sun2006,Holt2006,Regan2004}

The benzene molecules are farther from the CNT surface at the hot side than at the cold side, while the potential energy depends on the distance between them. In the temperature profile shown in Fig.~\ref{F3} (a), the central linear part follows Boltzmann statistics while the reservoirs at the endpoints are kept out of equilibrium by thermostats, in between there is a transition area where non-linearity presents. In Fig.~\ref{F3} (b), the force from the energy gradient is at the order of meV/nm, roughly in agreement with the pN value obtained from surface-free energy calculations.\cite{Panizon2017} Another force analysis in the Supporting Information (S3) shows that the force decreases to approximately zero when the system reaches equilibrium, where the mean thermophoretic driving force is canceled out by the scattering force in a steady state.

\begin{figure}[htp]
\centerline{\includegraphics[width=9cm]{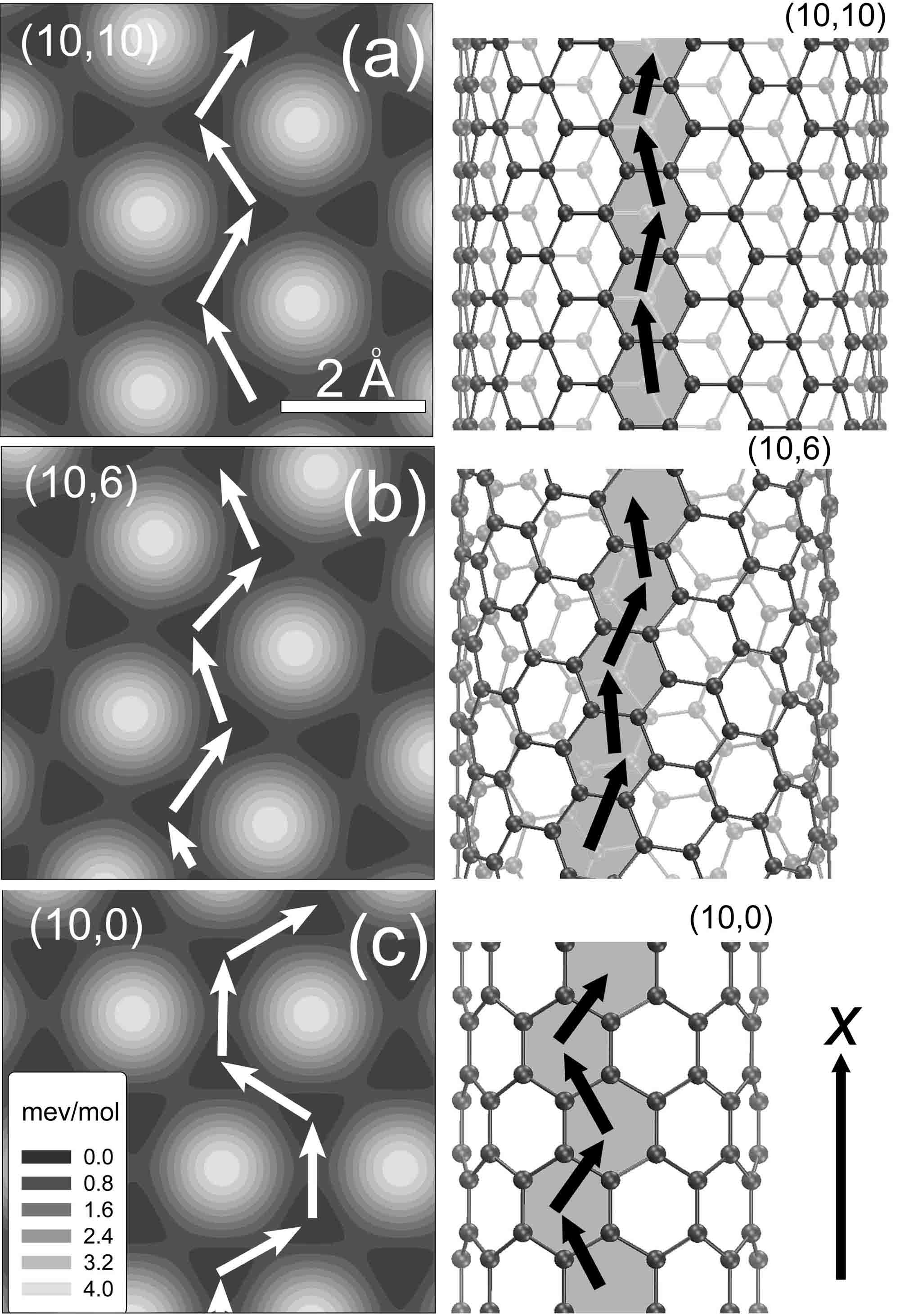}}
\caption{\label{F4}
Distribution of the vdW potential between a benzene molecule and (a) (10,10), (b) (10,6) and (c) (10,0) CNTs. The arrows indicate energetically optimized paths of the adsorbate. These energy landscapes are computed for a constant adsorbate-substrate distance of $0.34\;\mathrm{nm}$. Note that this calculation has also been performed for another distance $0.31\;\mathrm{nm}$, which results in similar potential landscapes with different energy corrugations.}
\end{figure}
 
The direction of this driving force exhibits strong dependence on the CNT structure at the local positions on the CNT. The energetically optimized path (EOP) on a potential energy surface (PES) indicates the most probable trajectory of an adsorbate on a substrate, according to the principle of minimizing energy corrugation.\cite{Wang2019a} i.e. the molecule will be less ``tired'' following an EOP on the CNT surface than any other trajectories. Figs.~\ref{F4} shows that a benzene molecule exhibits different EOPs on different CNTs. The key to the chirality dependence of the molecular transport on CNTs is that the free path for a different CNT is of a different length along $x$.

\begin{figure}[htp]
\centerline{\includegraphics[width=9cm]{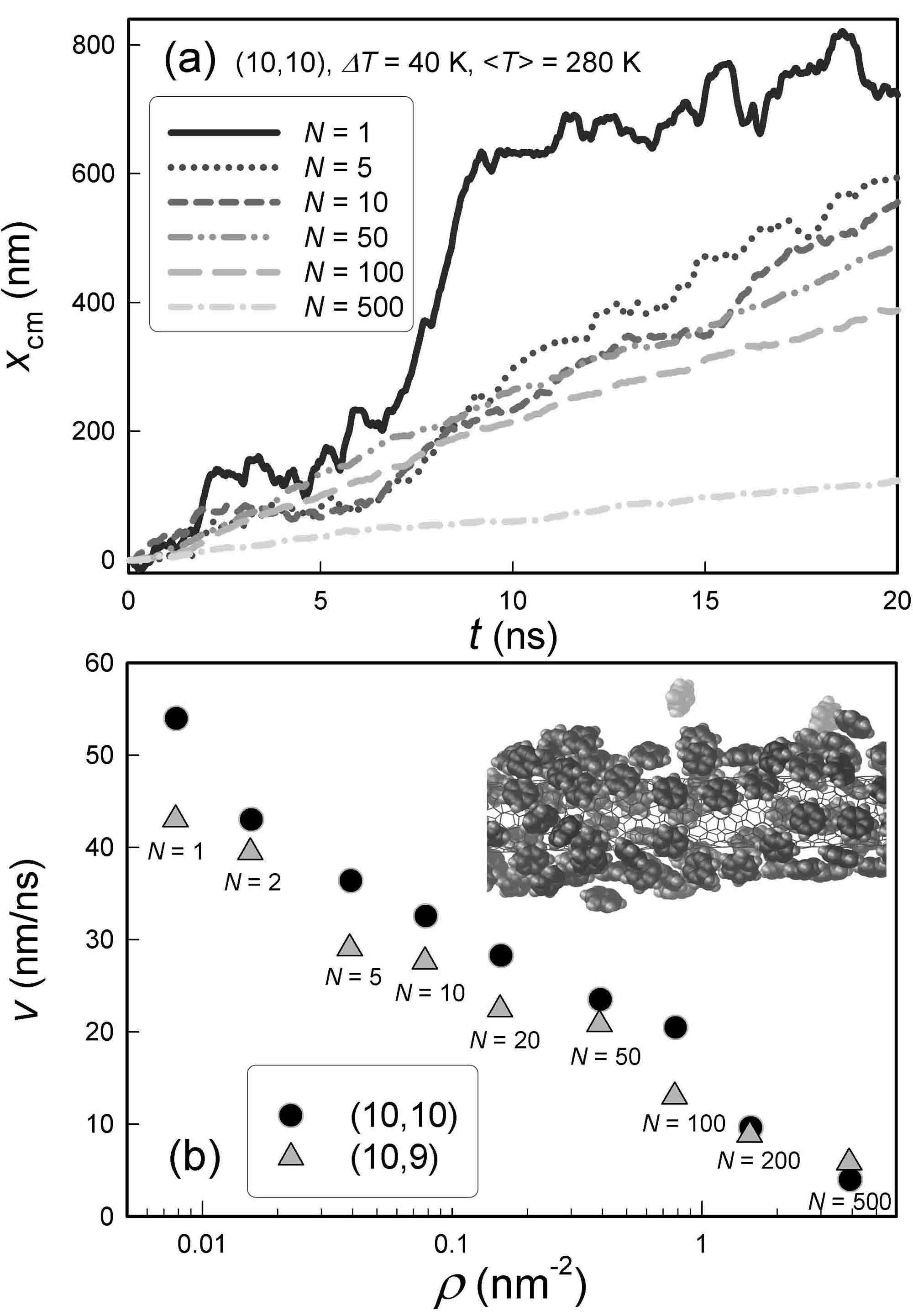}}
\caption{\label{F5}
(a) $x_{cm}$ vs $t$ for different numbers $N$ of benzene molecules adsorbed on a ($10$,$10$) CNT. (b) $v$ (slope of the $x_{cm}$ curve at $t>5\;\mathrm{ns}$) vs molecular densities $\rho=N/2\pi (R+0.34)p\ell_{0}$ for a ($10$,$10$) and a ($10$,$9$) CNTs. The inset shows a snapshot segment of $500$ benzene molecules drifting on a CNT.}
\end{figure}

A natural question to address is whether the molecular transport on armchair CNTs can be truly ballistic. This question leads to another set of simulations performed with various numbers $N$ of adsorbates on a ($10$,$10$) CNT. Fig.\ref{F5} (a) shows the displacement of the center of the mass of the molecules. It is seen a lower molecular density leads to a higher speed, because of a weaker probability of collision between molecules. A detailed analysis on the temperature distribution shows that the transport should be either ballistic or between ballistic and diffusive (namely quasi-ballistic), since the temperature profile of the molecules does not seem to be compatible with that of the CNT as shown in the Supporting Information (S4). 

The transport becomes very rapid for $N=1$ with a single molecule. It walked about $800\;\mathrm{nm}$ along the CNT in $20\;\mathrm{ns}$, by this speed it can travel through a centimeter-long CNT in a millisecond. Since it has no collision with other adsorbates, its $x_{cm}$ curve is expected to have a parabolic shape. However, this is not shown in Fig.\ref{F5} (a). The steady-state velocity may stem from possible collisions with thermally induced waves and ripples in the CNT surface (the ``friction'' of adsorbates) as well as other thermal effects including the random rotation of the molecules.\cite{Fasolino2007,Wang2011,Barreiro2008} Note that the oscillation of the $x_{cm}$ curves is observed to be more significant for small $N$ due to reduced sample population in statistics. 

Fig.\ref{F5} (b) provides a comprehensive view on the influence of molecular density $\rho$ on the transport. It is seen that $v$ increases monotonically with decreasing $\rho$. This is rather in contrast to thermophoresis of a gas mixture in which $v$ decreases with decreasing $\rho$ at low gas density. This contrast highlights the fact that the driving force comes mainly from the vdW interaction as shown in Fig.\ref{F3}. It is also shown that the drift velocity of the molecules on the ($10$,$10$) CNT is higher than that on the ($10$,$9$) one. The chirality effect remains even with very high or very low molecular density. Furthermore, dissociation of molecules can be observed for high molecular density, and as shown in a video recording in the Supporting Information (S9). The decrease in the drift velocity at high density may also be associated with the dissociation of the molecules from the CNT surface.

\begin{figure}[htp]
\centerline{\includegraphics[width=9.cm]{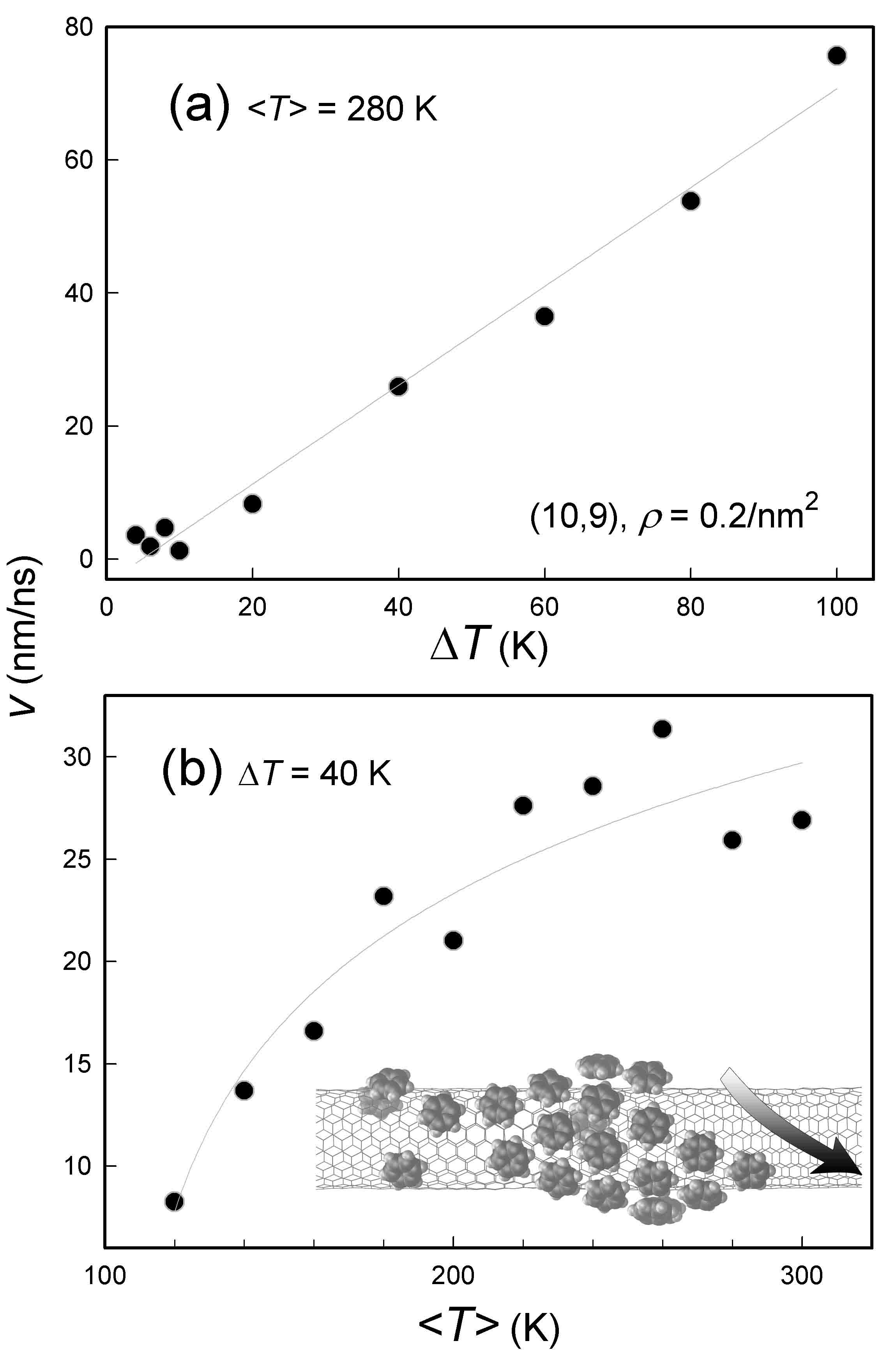}}
\caption{\label{F6}
(a) $v$ of $26$ benzene molecules on a ($10$,$9$) CNT vs the temperature difference $\Delta T=T_{1}-T_{2}$ at a given average temperature $\left\langle T \right\rangle=(T_{1}+T_{2})/2=280\;\mathrm{K}$. (b) $v$ at different $\left\langle T \right\rangle$ with a constant $\Delta T=20\;\mathrm{K}$. Inset shows a molecule aggregation formed on a CNT at low temperature. The eye-guiding curves reflect our interpretation of the data.}
\end{figure}

Aside from the chirality and molecular density, the temperature also plays an important role in the molecular transport. Different thermal gradients are used in the simulations as shown in Fig.\ref{F6} (a). These thermal gradients range from $0.2$ to $5.0\;\mathrm{K/nm}$, comparable to an experimentally-estimated thermal gradient of $1.0\;\mathrm{K/nm}$.\cite{Barreiro2008} It is seen that $v$ roughly holds a linear relationship with the temperature difference $\Delta T$. This implies that a constant drift mobility is kept, despite non-linear effects can be expected for large temperature gradients such as a $\Delta T$ of $100\;\mathrm{K}$ across a distance of $20\;\mathrm{nm}$. Regardless, this is consistent with the observation reported by Zambrano \textit{et al.} that the velocity of water nanodroplets inside CNTs is linearly proportional to the thermal gradient.\cite{Zambrano2009} The absence of the non-linear regime in the molecular conduction is in stark contrast with the conventional thermophoresis in gas or liquid phase and should be a unique feature on one-dimensional nanostructures. 

Fig.\ref{F6} (b) shows the influence of the average temperature on the molecular transport. No molecular drift is observable below an activation temperature of $100\;\mathrm{K}$, beyond which $v$ sub-linearly increases at increasing $\left\langle T \right\rangle$. This comes from a combination of temperature effects on both the molecular diffusivity and the driving force, since the friction force between the benzene and the graphene is expected to change with the velocity and to linearly increase at increasing temperature. At low pressure, the diffusion coefficient of gas molecules roughly increases with $T^{3/2}$,\cite{Cussler1997,Bartus2014} which can be approximated as a linear dependence for the here-used temperature range. The sub-linearly increase of $v$ is coupled with the fact that the anharmonicity of the vdW potential is enhanced at high temperature. 

Furthermore, the benzene molecules are interestingly found to form aggregates at low temperature $<150\;\mathrm{K}$. These aggregates is observed to move collectively following helical orbits on the CNT surface as shown in the inset of Fig.\ref{F6} (b) as well as in an animation in the Supporting Information (S7). This is consistent with the observation of Mao and Sinnott on the diffusive behavior of ethane and ethylene following a spiral path inside CNTs at low molecular densities,\cite{Mao2002} and is also correlated with the motion of a gold nanoparticle in CNTs.\cite{Schoen2006} The origin of the helical transitory of the molecules on the CNT surface can be explained by the dependence of the driving force direction on the CNT structure shown in Fig.\ref{F4}.

\section{Conclusions}

The chirality of CNT is found to play a central role in the molecular transport. The drift velocity of the molecules is found to be maximal on an armchair CNT, and to decrease with decreasing chiral angle. The chirality selectivity stems from a diversity of stacking configurations of $\pi$-orbitals at the interface. The driving force is attributed to a spatial variation in adsorbate-substrate interaction potential due to the non-uniform temperature field. This mechanism could be fundamental for explaining the undergoing physical mechanism of the molecular transport on the surface of nanostructures. 

The drift velocity is also found to increase with decreasing molecular density. The transport of less than fives molecules is identified to be ballistic or quasi-ballistic. A nonlinear regime of molecular transport is not detected since the drift mobility stays roughly constant even under very large thermal gradients. In contrast, the drift velocity increases with increasing temperature but shows a tendency to saturation, revealing a combined temperature effect on the molecular diffusivity and driving force.

\end{document}